\documentclass{ptephy_v1}

\preprintnumber{XXXX-XXXX} 

\usepackage{arydshln}

\begin{document}

\title{Determination of Trace Levels of Uranium and Thorium in High Purity Gadolinium Sulfate Using ICP-MS with Solid-Phase Chromatographic Extraction Resin}


\author{S. Ito}
\affil{Okayama University, Faculty of Science, Okayama 700-8530, Japan \email{s-ito@okayama-u.ac.jp}}

\author{Y. Takaku}
\affil{Institute for Environmental Sciences, Department of Radioecology, Aomori, 039-3212, Japan}

\author[3,4]{M. Ikeda}
\author[3,4]{Y. Kishimoto} 
\affil{Kamioka Observatory, Institute for Cosmic Ray Research, University of Tokyo, Kamioka, Gifu 506-1205, Japan}
\affil[4]{Kavli Institute for the Physics and Mathematics of the Universe (WPI), the University of Tokyo, Kashiwa, Chiba, 277-8582, Japan}



\begin{abstract}%

The new Super-Kamiokande-Gadolinium (SK-Gd) project is an upgrade of the Super-Kamiokande (SK) detector. 
In the SK-Gd project, 0.2\% Gd$_2$(SO$_4$)$_3$ is loaded into the 50 kton water tank of the SK. 
One of the main purposes of the project is to discover supernova relic neutrinos.  
Neutrino measurements and proton decay searches will also be performed in the SK-Gd. 
In order to measure solar neutrinos with a low energy threshold of $\sim$3.5 MeV in the SK-Gd, the main radioactive contaminations, $^{238}$U and $^{232}$Th, in Gd$_2$(SO$_4$)$_3{\cdot}$8H$_2$O, should be minimized before loading. 
Our maximum levels for U and Th are 5 mBq (U)/kg (Gd$_2$(SO$_4$)$_3{\cdot}$8H$_2$O) and 0.05 mBq (Th)/kg (Gd$_2$(SO$_4$)$_3{\cdot}$8H$_2$O). 

In order to measure such low concentrations of U and Th in Gd$_2$(SO$_4$)$_3{\cdot}$8H$_2$O, we developed the solid-phase extraction technique. 
Using this method, about 90\% or more U and Th could be efficiently extracted while  Gd was reduced by a factor of about $10^{4}$. 
This allowed these radioactivity contaminations to be measured  precisely as 0.04 mBq/kg (Gd$_2$(SO$_4$)$_3{\cdot}$8H$_2$O) for U and 0.01 mBq/kg (Gd$_2$(SO$_4$)$_3{\cdot}$8H$_2$O) for Th. 
We measured three pure Gd$_2$(SO$_4$)$_3{\cdot}$8H$_2$O samples using this method and estimated that the purest one contained $<0.04$ mBq (U)/kg (Gd$_2$(SO$_4$)$_3{\cdot}$8H$_2$O) and 0.06 $\pm$ 0.01 mBq (Th)/kg (Gd$_2$(SO$_4$)$_3{\cdot}$8H$_2$O) by ICP-MS. 

\end{abstract}

\subjectindex{The Super-Kamiokande, ICP-MS, Solid-phase extraction}

\maketitle

\section{Introduction}

\subsection{The Super-Kamiokande and the new SK-Gd project}
The Super-Kamiokande, which is located 1000 m underground at Ikenoyama, Kamioka, Gifu, Japan, is the world's largest water Cherenkov detector with a mass of 50 kton \cite{NIMA}. 
The main physics targets for the SK are proton decay searches and measuring neutrinos from various sources such as the Sun, the atmosphere, supernova explosions, and artificial neutrino beams. 
The energy threshold is set at about 3.5 MeV to measure low energy neutrinos, such as solar neutrinos.

On 27th June 2015, the Super-Kamiokand-Gadolinium (SK-Gd) project was approved by the SK Collaboration; this is an upgrade of the SK detector, in which 0.2\% gadolinium sulfate (Gd$_2$(SO$_4$)$_3$) is added to the SK water tank \cite{Gd}. 
One of the goals of the SK-Gd is to discover supernova relic neutrinos and study the star formation history of the universe \cite{SRN}.
Gadolinium sulfate octahydrate (Gd$_2$(SO$_4$)$_3{\cdot}$8H$_2$O) is used for the SK-Gd, as it has higher solubility than unhydrated gadolinium sulfate. 
We aim to introduce Gd$_2$(SO$_4$)$_3{\cdot}$8H$_2$O in JFY-2019 and then start the SK-Gd experiment.

The concentration of radioactive impurities in Gd$_2$(SO$_4$)$_3{\cdot}$8H$_2$O loaded water needs to be kept at the same level as in the current SK water in order to continue precise measurements of the solar neutrino spectrum \cite{solar}. 
Thus, Gd$_2$(SO$_4$)$_3{\cdot}$8H$_2$O should be purified before loading into the SK water tank to avoid interference with the solar neutrino measurements. 
The maximum levels for the main radioactive contaminations, U and Th, are 5 mBq (U)/kg (Gd$_2$(SO$_4$)$_3{\cdot}$8H$_2$O) and 0.05 mBq (Th)/kg  (Gd$_2$(SO$_4$)$_3{\cdot}$8H$_2$O)\footnote{Unless otherwise specified, the unit ``mBq/kg"  represents mBq of U or Th in kg of Gd$_2$(SO$_4$)$_3{\cdot}$8H$_2$O powder.}, respectively.

\subsection{ICP-MS measurement}

Inductively Coupled Plasma-Mass Spectrometry (ICP-MS) is a highly sensitive mass spectrometry technique that is widely used in analytical chemistry to measure elements in solution. 
As $^{238}$U and $^{232}$Th have long half-lives (4.5${\times}10^{9}$ and 1.4${\times}10^{10}$ years, respectively), ICP-MS has a higher sensitivity to their radioactivity than other detectors (e.g., Ge detectors). 
Introducing highly concentrated solution into the ICP-MS, however, affects the sensitivity. 
High matrix elements interfere with the ionization of target elements, which reduces the sensitivity of the ICP-MS. 
This problem is generally called ``matrix effect".

The introduction of high matrix elements into the ICP-MS also increases the number of polyatomic ions. 
If many polyatomic ions originating from high matrix elements (in our case, Gd and O in the water, and Ar, which is used for Ar plasma in the ICP-MS)  are produced, whose mass numbers are similar to those of target elements (in our case, $^{238}$U and $^{232}$Th) in the ICP-MS, the target elements would be overestimated. 
This issue is called ``spectral interference". 
In our case, potential spectral interference resulting from Gd, O, and Ar would be caused by $^{158}$Gd$^{40}$Ar$_2$, $^{158}$Gd$^{16}$O$_5$, and so on,  for $^{238}$U, and $^{152}$Gd$^{40}$Ar$_2$, $^{160}$Gd$^{18}$O$_4$, and so on, for $^{232}$Th.

In order to remove this interference, matrix elements should be reduced to at least $<100~{\mu}$g mL$^{-1}$. 
It is, however, difficult to measure U and Th with the high precision that we require by the dilution of Gd$_2$(SO$_4$)$_3{\cdot}$8H$_2$O solution. 
Therefore, we developed a solid-phase extraction method to extract U and Th from Gd$_2$(SO$_4$)$_3{\cdot}$8H$_2$O and measured these concentrations  precisely at an order of 0.01 mBq/kg using ICP-MS. 
We report on this method in this paper.

\section{Experimental equipment}

\subsection{ICP-MS in Kamioka}

\begin{figure}[htbp]
\centerline{\includegraphics[width=14cm]{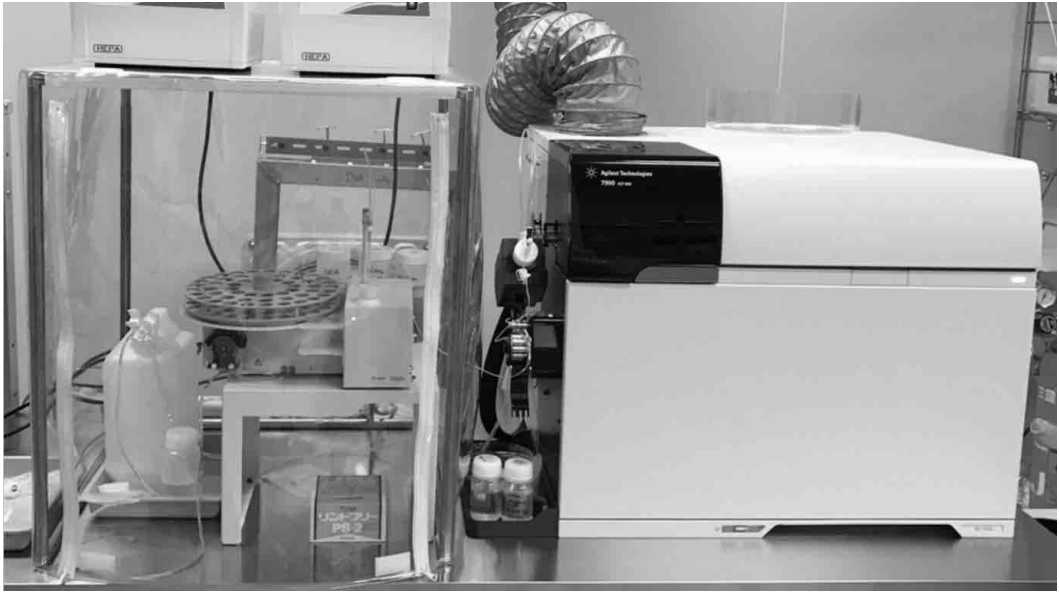}}
\caption{Picture of the ICP-MS ``Agilent7900" in the Kamioka clean room. The auto-sampler is fully covered by the class 100 clean booth to reduce contamination from the outside.}
\label{fig:Agilent7900}
\end{figure}

\begin{figure}[htbp]
\centerline{\includegraphics[width=15cm]{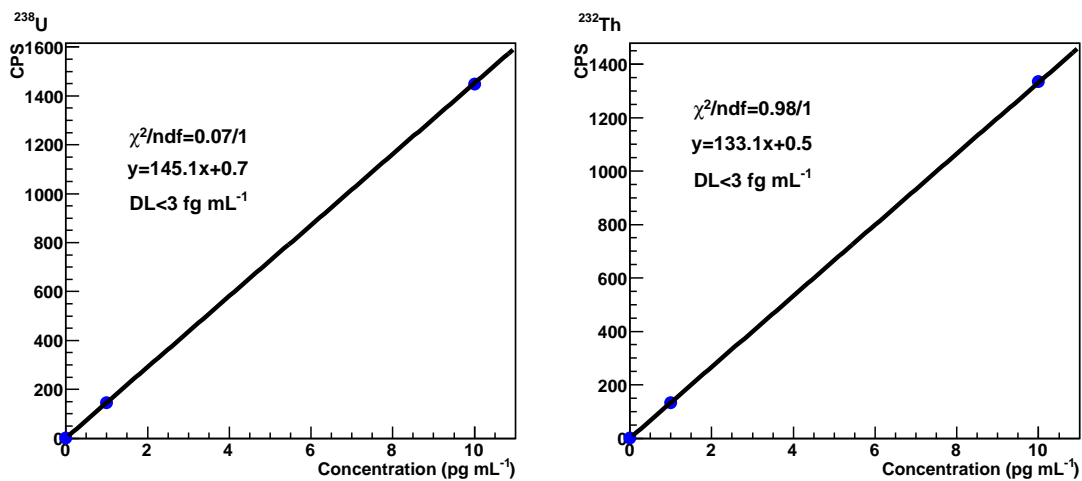}}
\caption{Calibration curves of $^{238}$U (left) and $^{232}$Th (right) with He gas. The horizontal axes show concentrations of U and Th, and the vertical axes represent counts per second (CPS). These points are fitted to linear functions.}
\label{fig:Calib}
\end{figure}

In December 2016, an ICP-MS ``Agilent 7900" (Agilent Technologies \cite{Agilent}) was installed in the Kamioka clean room, a clean room in the Kamioka mine near the SK detector. 
Figure \ref{fig:Agilent7900} shows a picture of the ICP-MS in the Kamioka clean room. 
The main characteristic of the Agilent 7900 is highly sensitive ICP-MS because of low dark noise. 
Additionally, the collision/reaction cell of the ICP-MS is filled with helium (He) gas in order to reduce dark counts originating from polyatomic ions. 
When He gas is not used, the dark counts from polyatomic ions at mass numbers of 232 (Th) and 238 (U) in 0.2 mol L$^{-1}$ nitric acid (HNO$_3$) blank solution are both about 10 counts per second (CPS). 
When using He gas, the counts of polyatomic ions could be reduced to $<1$ CPS. 
Figure \ref{fig:Calib} shows the calibration curves of U and Th for the ICP-MS. 
The detection limit (DL) of the ICP-MS can be defined as
\begin{eqnarray}
{\rm DL}&<&\frac{3{\sigma}_{\rm blank}}{a},
\end{eqnarray}
where ${\sigma}_{\rm blank}$ and $a$ represent the statistical uncertainty of the CPS for the blank solution and the slope of the calibration curve with He gas, respectively. 
While the values of DL for U and Th were respectively 13 fg mL$^{-1}$ and 15 fg mL$^{-1}$ without He gas, this improved to 3 fg mL$^{-1}$ when He gas was introduced.

\begin{figure}[htbp]
\centerline{\includegraphics[width=12cm]{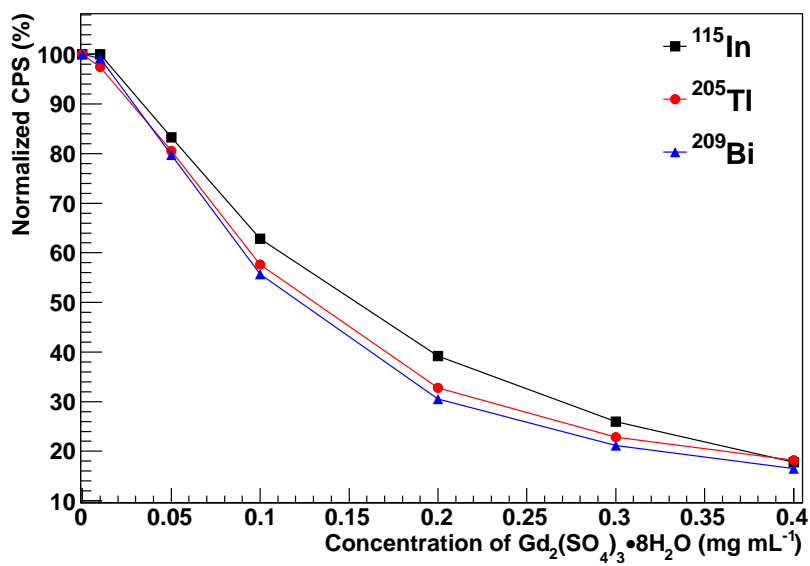}}
\caption{Typical example of the matrix effect on Gd$_2$(SO$_4$)$_3{\cdot}$8H$_2$O. The horizontal axis shows the concentrations of Gd$_2$(SO$_4$)$_3{\cdot}$8H$_2$O (mg mL$^{-1}$) and the vertical axis represents the normalized counts per second (CPS). The CPS for $^{115}$In, $^{205}$Tl, and $^{209}$Bi were normalized to the CPS for each element in the blank solution. }
\label{fig:ISTD}
\end{figure}

Figure \ref{fig:ISTD} shows a typical example of the matrix effect from various concentrations of  Gd$_2$(SO$_4$)$_3{\cdot}$8H$_2$O on the sensitivity of the ICP-MS. 
For this study, 10 ng mL$^{-1}$ of $^{115}$In, $^{205}$Tl, and $^{209}$Bi mixed solution was used. 
Comapred with the blank solution, the CPS of all elements dropped to 30$\sim$40\% at 0.2 mg mL$^{-1}$ Gd$_2$(SO$_4$)$_3{\cdot}$8H$_2$O solution due to the matrix effect of Gd.  

The Kamioka clean room was designed as class 1000, and the auto-sampler is fully covered by the clean booth, which is designed as class 100. 
This clean environment is setup to minimize contamination from the outside.

\subsection{Reagents and equipment}

\begin{figure}[htbp]
\centerline{\includegraphics[width=8cm]{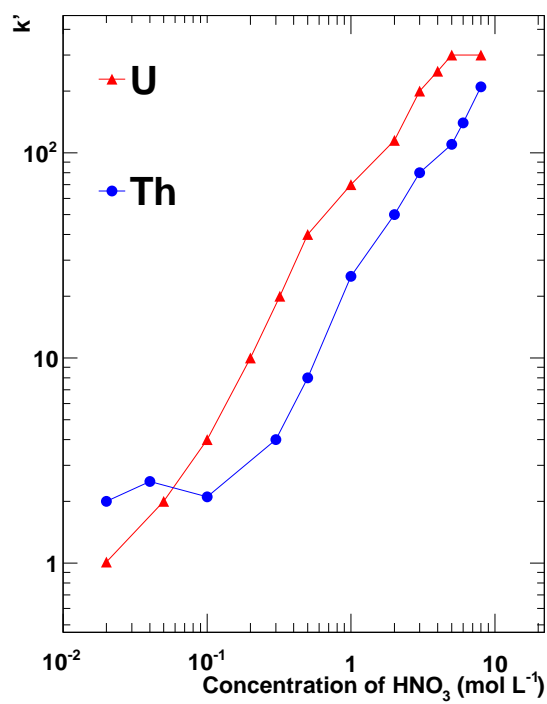}}
\caption{HNO$_3$ concentration dependence for U (red triangles) and Th (blue circles) with the UTEVA resin \cite{Howritz}. The horizontal axis shows the  concentration of HNO$_3$ and the vertical axis shows the distribution coefficient $(k')$.}
\label{fig:UTEVA}
\end{figure}

\paragraph{Chromatographic extraction resin}
In this study, we used UTEVA resin, which is a commercially available product of Eichrom Technologies LLC \cite{UTEVA} and has been widely used for a variety of analytical separations. 
Figure \ref{fig:UTEVA} shows the HNO$_3$ concentration dependence for U and Th with the UTEVA resin \cite{Howritz}. 
All the tetravalent actinides and U(V\hspace{-.1em}I) have strong retention (distribution coefficient $k'>100$, which is defined as the ratio of the number of chemicals (U and Th) in the resin to the number of the chemicals in the solution \cite{UTEVA, Howritz}) with the UTEVA resin in $>$5 mol L$^{-1}$ HNO$_3$. 
On the other hand, Gd forms a trivalent ion so that Th and U can be extracted from Gd$_2$(SO$_4$)$_3{\cdot}$8H$_2$O using the UTEVA resin. 
These features make it possible to separate U and Th from Gd-loaded water.

\paragraph{Reagents}
In order to produce solutions with low contamination, the ultra-pure SK water \cite{NIMA} and ultra-high-purity analytical grade 68\% HNO$_3$ (TAMAPURE AA-100 made by Tama Chemicals \cite{Tama}) were used. 
Calibration of the ICP-MS and estimation of the recovery for U and Th were performed using a high-purity 29 element mixed standard solution (XSTC-331, SPEX Inc.  \cite{spex}). 
In order to check the magnitude of the matrix effects, In, Tl, and Bi standard solutions (Kanto Chemical Co., Inc. \cite{Kanto}) were used as internal standard solutions. 
These solutions were mixed and diluted to 10 ng mL$^{-1}$ in 0.2 mol L$^{-1}$ HNO$_3$ solution. 
Electronic (EL) grades of 70\% HNO$_3$ and 35\% hydrochloric acid (HCl) (Wako Pure Chemical Industries, Ltd. \cite{Wako}) were used to prewash the resin, column, bottles, and other equipment.

\begin{figure}[htbp]
\centerline{\includegraphics[width=9.5cm, trim= 8cm 6cm 8cm 5cm, clip]{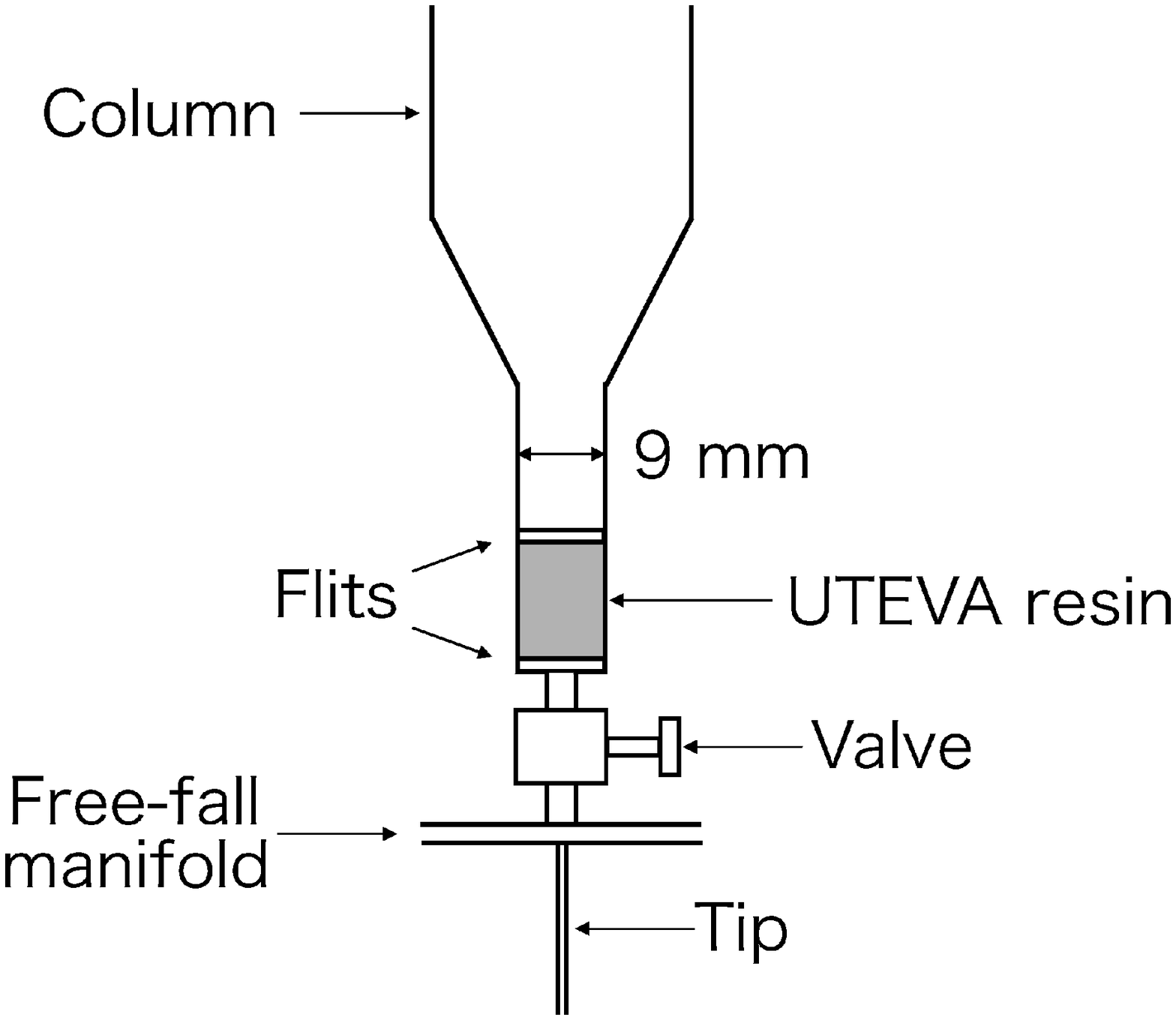}}
\caption{Schematic of the experimental setup.}
\label{fig:column}
\end{figure}

\paragraph{Column and equipment for solid-phase extraction}
A column with a volume of about 20mL and an inside diameter of 9 mm (LSC-$\phi$9, GL Sciences Inc. \cite{GLS}) was used for this study. 
Figure \ref{fig:column} shows a schematic of the experimental setup. 
A valve was added to the column in order to adjust the flow rate. 
This setup was connected to a tip, and fixed to a free-fall manifold. 
The UTEVA resin was sandwiched between two flits to fix.

\paragraph{Bottles}
All bottles used for this study were made of polypropylene. 
Bottles with a volume of 100 mL were used for the preservation of the sample solutions and the collection of solutions during the solid-phase extraction procedure. 
Bottles with a volume of 14 mL were used for the ICP-MS measurements to place samples on the auto-sampler.

\section{Study of the solid-phase extraction}

\subsection{Procedure}

\begin{figure}[htbp]
\vspace{4mm}
\centerline{\includegraphics[width=15cm]{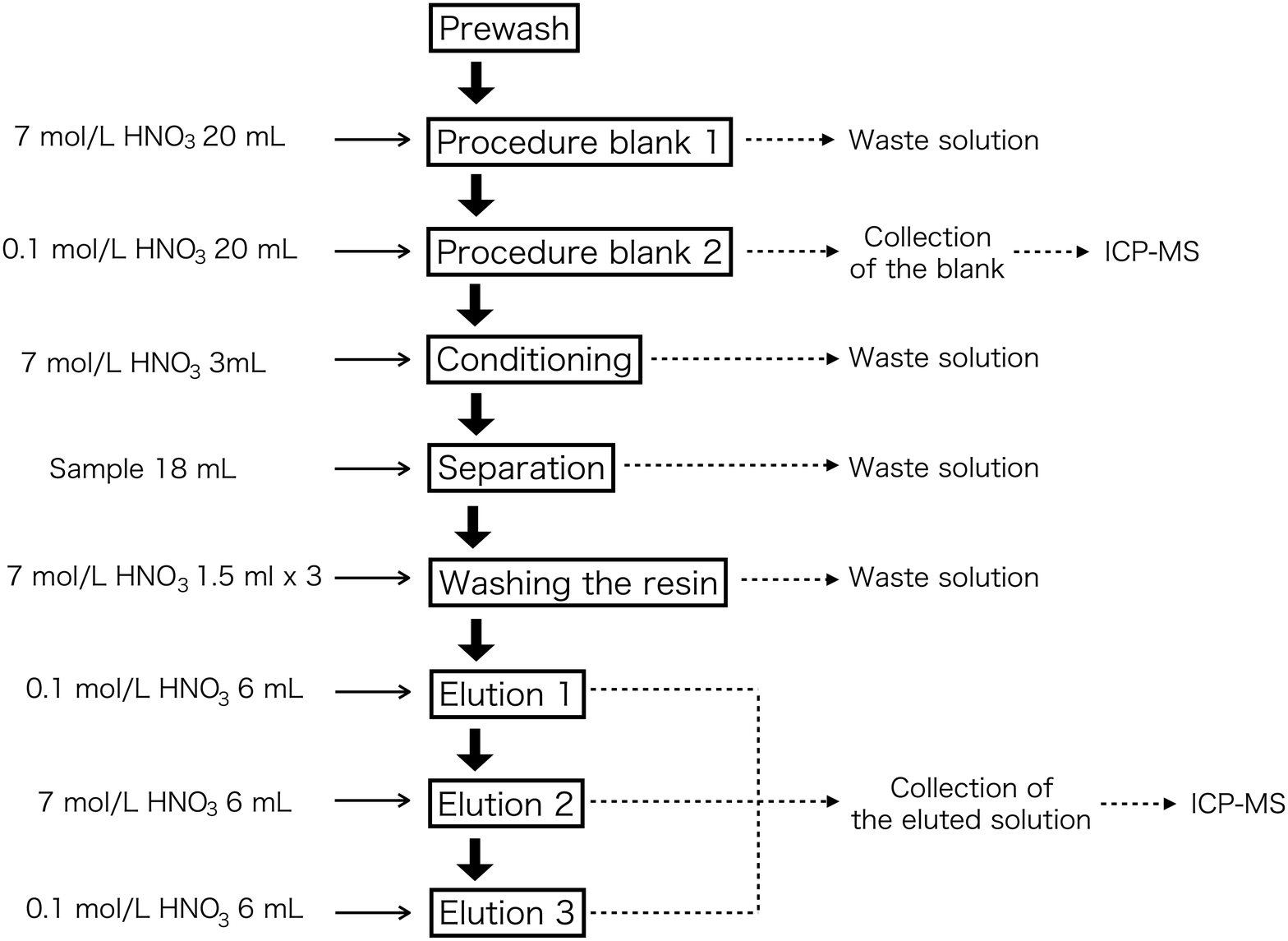}}
\caption{Diagram of the whole procedure for the solid-phase extraction.}
\label{fig:diagram}
\end{figure}

The specially purified Gd$_2$(SO$_4$)$_3{\cdot}$8H$_2$O samples were provided by various  companies; we call them samples A, B, and C in this paper.\footnote{For the contract between companies and us, the companies' and products' names are blinded.}  
Figure \ref{fig:diagram} shows a diagram of the whole procedure. 
The details of each step are described in this section.

\subsubsection{Prewash}

The resin and equipment were suspected to be contaminated during the production process. 
Thus, the following prewash procedures were performed. 
The UTEVA resin was soaked in 0.5 mol L$^{-1}$ HNO$_3$ solution (EL grade) for more than one night. 
The columns, bottles, and all other equipment were also soaked in 1 mol L$^{-1}$ of HNO$_3$ solution (EL grade) for at least one night. 
All of the soaked items were washed with the ultra-pure SK water. 
Then, the column was charged with 0.3 g of the UTEVA ($\sim$1.5 cm of the bed height) and the column, the valve, and the tip were set up as illustrated in Figure \ref{fig:column}. 
The resin and inside of the column, the valve, and the tip were washed with 20 mL of 7 mol L$^{-1}$ HNO$_3$ (EL grade) followed by 20 mL of 1 mol L$^{-1}$ HCl (EL grade). 
Finally, this setup was washed by loading 20 mL of ultra-pure SK water. 
In the prewash procedure, the valve was fully opened (the flow rate was about 5 mL/min). 
The temperature of all the solutions was the same with the temperature in the clean room being about 22 $^{\circ}$C.

\subsubsection{Procedure blank}

In order to collect the procedure blank solution, 20 mL of 7 mol L$^{-1}$ HNO$_3$ solution was loaded into the column. 
After all the solution passed through the column, 20 mL of 0.1 mol L$^{-1}$ HNO$_3$ solution was loaded into it. 
Then, the 0.1 mol L$^{-1}$ HNO$_3$ solution that passed through the column was collected as the procedure blank solution. 
If the prewash procedure was skipped, more than 1 pg mL$^{-1}$ of U and Th contamination was observed in the procedure blank solution. 
On the other hand, the concentration of U and Th in the procedure blank solution could be reduced to about 10 fg mL$^{-1}$ after the prewash procedure.  

\subsubsection{Conditioning}

To condition the resin, 3 mL of 7 mol L$^{-1}$ HNO$_3$ solution was loaded into the column. 
During this procedure, the valve was adjusted to set the flow rate to about 0.3 mL/min in order to efficiently adsorb U and Th in the resin for the next procedure.

\subsubsection{Separation}

For this study, 1.0 g of the new Gd$_2$(SO$_4$)$_3{\cdot}$8H$_2$O sample was dissolved in 100 mL of 7 mol L$^{-1}$ HNO$_3$ solution. 
In order to separate U and Th from the Gd$_2$(SO$_4$)$_3{\cdot}$8H$_2$O solution, 18 mL of the sample solution was loaded into the column with a flow rate of about 0.3 mL/min. 
The valve was fully opened again after all the sample solution had passed through.

\subsubsection{Washing the resin}

Before eluting the U and Th, the resin should be washed in order to reduce the remaining Gd contained in it. 
The resin was washed three times with 1.5 mL of 7 mol L$^{-1}$ HNO$_3$. 
When this procedure was not performed, the count rate of the ICP-MS decreased to about 50\% due to the matrix effect from the remaining Gd. 
According to Figure \ref{fig:ISTD}, about 10\% of the Gd remained in the resin before this procedure.

\subsubsection{Elution}

In general, U and Th in the UTEVA resin can be eluted with dilute HNO$_3$ solution. 
However, U and Th are adsorbed by several materials in low concentration acid solution \cite{Book, Th}. 
In particular, Th is adsorbed to a greater extent than U and this effect is significant for $<$1 pg mL$^{-1}$ level measurements. 
In order to reduce such a problem, the elution procedure was improved as follows: 6 mL of 0.1 mol L$^{-1}$ HNO$_3$ was loaded into the column to break the chemical bonds between the resin and U/Th (more than 80\% of U is eluted in this procedure). 
Next, 6 mL of 7 mol L$^{-1}$  HNO$_3$ was loaded to remove and collect Th on the columns or the surface of the resin. 
Finally, 6 mL of 0.1 mol L$^{-1}$ HNO$_3$ was loaded to collect the remaining U and Th. 
These eluted solutions were collected in the same bottle (18 mL in total) and measured using ICP-MS. 

\subsection{Estimation of recovery rate}

\subsubsection{Extraction of U and Th from non-Gd solution}\label{sec:recovery}

\begin{figure}[htbp]
\centerline{\includegraphics[width=12cm]{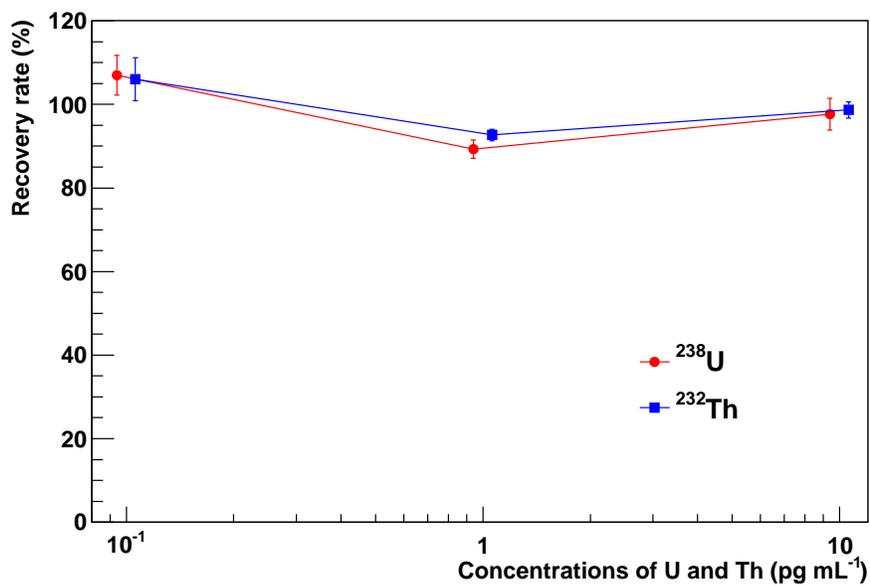}}
\caption{Recovery rates of the eluted solution for U (red circles) and Th (blue squares). The horizontal axis shows the concentrations of U and Th, and the  vertical axis represents the recovery rate. }
\label{fig:Recovery}
\end{figure}

In order to estimate the recovery rate (the U and Th recovered in this study divided by added U and Th), 7 mol L$^{-1}$ HNO$_3$ solutions with U and Th with  no Gd$_2$(SO$_4$)$_3{\cdot}$8H$_2$O were prepared. 
The concentration of U and Th in the sample solutions was adjusted to 0.1, 1, and 10 pg mL$^{-1}$. 
Figure \ref{fig:Recovery} shows the results of this study. 
By this method, about 90\% or more of the U and Th could be recovered. 

\subsubsection{Extraction of U and Th from Gd$_2$(SO$_4$)$_3{\cdot}$8H$_2$O solution}

\begin{figure}[htbp]
\centerline{\includegraphics[width=15cm]{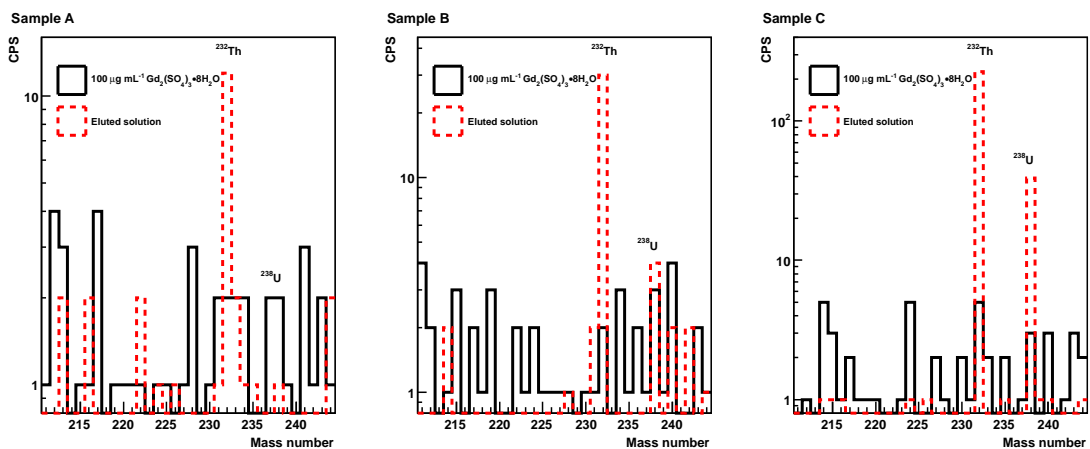}}
\caption{Mass spectra of the eluted solution (dashed red) and 100 $\mu$g mL$^{-1}$ of Gd$_2$(SO$_4$)$_3{\cdot}$8H$_2$O solutions of sample A (left), B (center), and C (right) in the mass range 211 to 244.}
\label{fig:Spectrum4}
\end{figure}

\begin{figure}[htbp]
\centerline{\includegraphics[width=15cm]{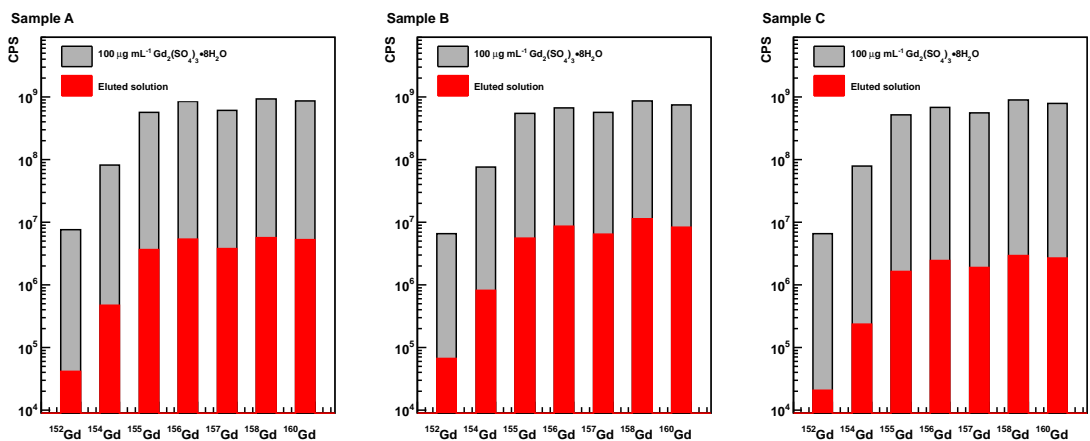}}
\caption{CPS of all Gd isotopes for samples A (left), B (center), and C (right) before and after chemical extraction. The gray histograms show the 100 $\mu$g mL$^{-1}$  Gd$_2$(SO$_4$)$_3{\cdot}$8H$_2$O solutions and the red histograms indicate the CPS of Gd in the eluted solutions.}
\label{fig:GdSpectrum}
\end{figure}

\paragraph{Spectral interference}
The effects of the spectral interference from Gd and the efficiency of the separation between U, Th, and Gd were studied.  
For these studies, 1.00 g of the three new Gd$_2$(SO$_4$)$_3{\cdot}$8H$_2$O samples A, B, and C was dissolved in 100 mL of 7 mol L$^{-1}$ HNO$_3$ and the developed solid-phase extractions were performed. 
Figure \ref{fig:Spectrum4} shows mass spectra of the three eluted solutions for a mass range of 211 to 244. 
For reference, the mass spectra of Gd$_2$(SO$_4$)$_3{\cdot}$8H$_2$O solutions are also shown. 
In order to the reduce matrix effect, the Gd$_2$(SO$_4$)$_3{\cdot}$8H$_2$O solutions were diluted to 100 $\mu$g mL$^{-1}$. 
Compared to the reference spectra, we can see that the spectral interference was efficiently reduced by our separation method.  
Figure \ref{fig:GdSpectrum} shows mass spectra of Gd in the eluted solutions and 100 $\mu$g mL$^{-1}$ Gd$_2$(SO$_4$)$_3{\cdot}$8H$_2$O solutions for reference. 
Because the Gd spectra for all the eluted solutions could be reduced to an order of $10^{-6}$ g mL$^{-1}$, the effects of the spectral interference on U and Th were reduced to a negligible level.

\begin{figure}[htbp]
\centerline{\includegraphics[width=15cm]{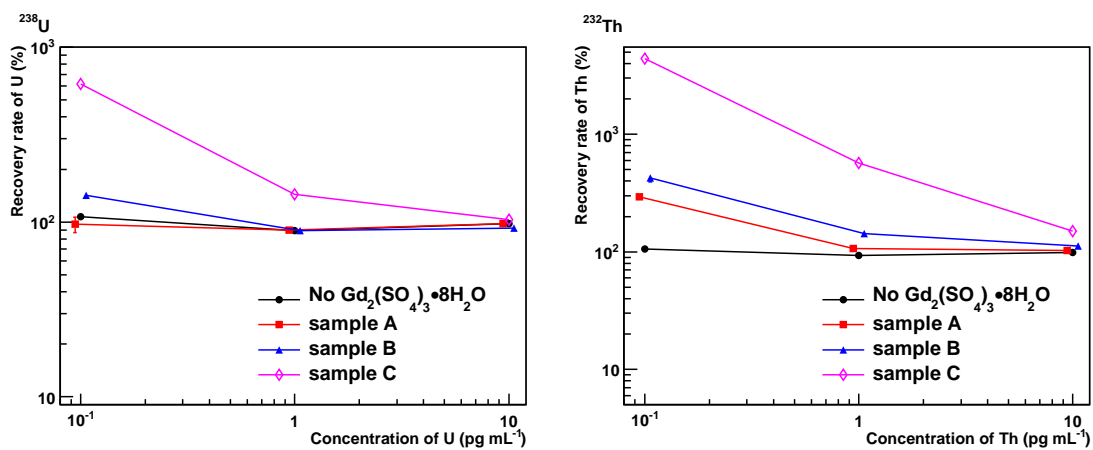}}
\caption{Recovery rates of U (left) and Th (right) with and without Gd$_2$(SO$_4$)$_3{\cdot}$8H$_2$O samples. The horizontal axes show the concentrations  of U and Th, and vertical axes represent the recovery rates for U and Th. Black circles, red squares, blue triangles, and pink diamonds indicate the sample without Gd$_2$(SO$_4$)$_3{\cdot}$8H$_2$O sample, and samples A, B, and C, respectively.}
\label{fig:Recovery2}
\end{figure}

\paragraph{Recovery rate}
The influence of Gd$_2$(SO$_4$)$_3{\cdot}$8H$_2$O on the retention of the resin was also studied by adding U and Th to the sample solutions with the concentration of 0.1, 1, and 10 pg mL$^{-1}$. 
Figure \ref{fig:Recovery2} shows the results of this study. 
We can see that the recovery rates were remained at $>~90$\% for all samples at all concentrations of U and Th. 
Because of U and Th impurities in the Gd$_2$(SO$_4$)$_3{\cdot}$8H$_2$O samples, the recovery rates of U and Th at 0.1 and 1 pg mL$^{-1}$, especially sample C highly exceed 100\%. 
The measured concentrations of U and Th will be described in the next section in more detail. 
In this study, we confirmed that Gd$_2$(SO$_4$)$_3{\cdot}$8H$_2$O did not influence the recovery rate.

\begin{figure}[htbp]
\centerline{\includegraphics[width=12cm]{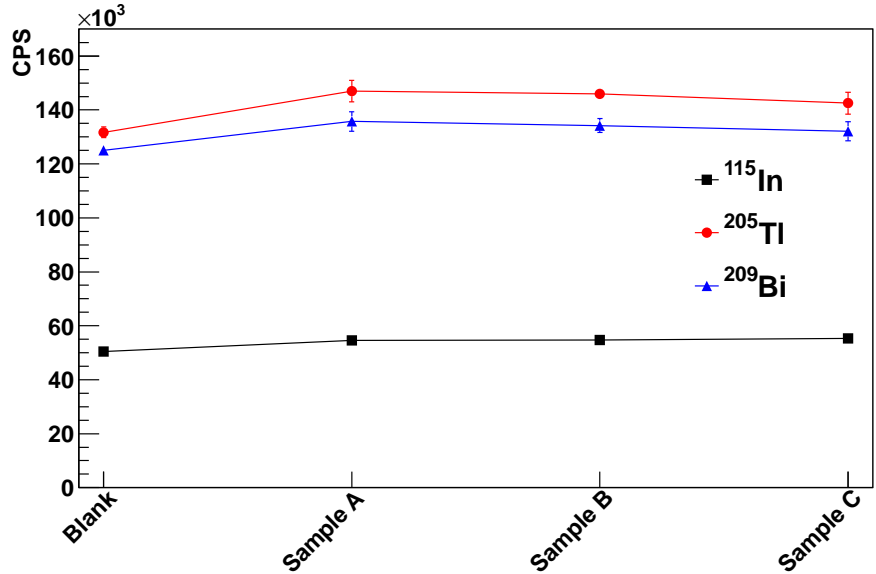}}
\caption{Variation of the count rate for the internal standard solution, In (black squares), Tl (red circles), and Bi (blue triangles) for the blank and eluted solutions of samples A, B, and C. Since the mass number of In is about half that of Tl and Bi, the count rate of In is also different to the others.}
\label{fig:ISTD2}
\end{figure}

\paragraph{Matrix effect}
During the ICP-MS measurements, the variation in the sensitivity of this technique was always monitored by mixing the internal standard solution (10 ng mL$^{-1}$ In, Tl, and Bi) into the sample solutions. 
Figure \ref{fig:ISTD2} shows the variation in the internal standard solution for all samples. 
In this study, no decrease in the sensitivity of ICP-MS due to the matrix effect from Gd was observed.

\section{Results and discussion}

The results of the measurements of U and Th in samples A, B, and C are listed in Table \ref{tab:results}. 
For comparison, the values for the procedure blank solution, typical commercially available Gd$_2$(SO$_4$)$_3{\cdot}$8H$_2$O, and our requirements are also given. 
The relationships between pg g$^{-1}$ (Gd$_2$(SO$_4$)$_3{\cdot}$8H$_2$O) and mBq/kg for U and Th are 
\begin{eqnarray}
1~{\rm mBq~(U)/kg}&=&81.3~{\rm pg~(U)~g^{-1}}~{\rm and}\\
1~{\rm mBq~(Th)/kg}&=&246~{\rm pg~(Th)~g^{-1}}.
\end{eqnarray}
As shown in Figure \ref{fig:Recovery}, the recovery rates for U and Th at 0.1 pg mL$^{-1}$ are  (107 $\pm$ 5)\% and (106 $\pm$ 5)\%,  respectively, while  the recovery rates at 1 pg mL$^{-1}$ for U and Th are respectively (89 $\pm$ 2)\% and (93 $\pm$ 1)\%. 
Therefore, 15\% uncertainty was conservatively added to the conversion from the solution to the powder.

The DLs of U and Th with this developed method were limited by the procedure blank solution and the uncertainty in the recovery rate since the matrix effect and spectrum interference were reduced to a negligible level. 
The concentration of U and Th in the procedure blank solution were about 0.01 pg mL$^{-1}$. 
By conservatively taking three times these values inflated by 15\%, which is the uncertainty in the recovery rate, the DLs of U and Th were estimated to be 0.04 mBq (U)/kg and 0.01 mBq (Th)/kg. 
These values are lower than our requirements by two orders of magnitude for U and a factor of five for Th.

By the efforts of various companies, U was well purified about one or two  orders of magnitude lower than our requirement. 
With respect to Th contamination, sample A almost reaches our goal, sample B should be further purified by a factor of 2.4, and sample C showed two orders of magnitude larger than the required concentration. 
At present, all companies make purer Gd$_2$(SO$_4$)$_{3}{\cdot}$8H$_2$O. 
We expect to obtain ultra-pure Gd$_2$(SO$_4$)$_{3}{\cdot}$8H$_2$O soon and load it into the SK in the near future.

\begin{table}[htbp]
\begin{center}
\caption{Results for samples A, B, C, and the values for the procedure blank solution, a typical commercially available product, and our requirements. The values for U and Th in the procedure blank are already subtracted from the values for U and Th for all the samples. The value of U for sample A was estimated using 3$\sigma$ with a 15\% inflation of the ICP-MS measurements.}
\tabcolsep7pt\begin{tabular}{c|cc}\hline
 & U (pg mL$^{-1}$) & U (mBq/kg)  \\ 
 & ICP-MS measurement & Converted to powder  \\ \hline
Sample A & 0.02 $\pm$ 0.01 & $<~0.04$ \\
Sample B & 0.04 $\pm$ 0.01 & 0.05 $\pm$ 0.01 \\
Sample C & 0.52 $\pm$ 0.01 & 0.64 $\pm$ 0.10 \\
Procedure blank & 0.01 & - \\
Commercial\footnotemark[3] & 0.23 & 28$\pm$2 \\ \hdashline
Requirement & - & 5 \\ \hline \hline
& Th (pg mL$^{-1}$) & Th (mBq/kg) \\
& ICP-MS measurement & Converted to powder\\ \hline
Sample A& 0.15 $\pm$ 0.01 & 0.06 $\pm$ 0.01\\
Sample B& 0.30 $\pm$ 0.01 & 0.12 $\pm$ 0.02\\
Sample C& 4.35 $\pm$ 0.18 & 1.77 $\pm$ 0.27\\
Procedure blank & 0.01 & - \\
Commercial\footnotemark[3] & 0.47 & 19 $\pm$ 1 \\ \hdashline
Requirement & - & 0.05\\ \hline
  \end{tabular}
\label{tab:results}
\end{center}
\end{table}
\footnotetext[3]{The concentrations of U and Th are much enough to measure without solid-phase extraction method. Therefore, 100 $\mu$g mL$^{-1}$ of Gd$_2$(SO$_4$)$_{3}{\cdot}$8H$_2$O solution was directly measured by the ICP-MS.}

\section{Conclusions}

A new SK-Gd project that involves Gd$_2$(SO$_4$)$_{3}{\cdot}$8H$_2$O into the SK water tank is in preparation. 
In order to measure radioactive impurities in Gd$_2$(SO$_4$)$_{3}{\cdot}$8H$_2$O, we developed a solid-phase extraction method using chromatographic extraction resin. 
This method has a high recovery rate for U and Th of about 90\% or more with a high efficiency of separation of U and Th from Gd of a factor of about $10^4$. 
The matrix effect and spectral interference in the ICP-MS measurements were effectively reduced and we could measure U and Th in three pure Gd$_2$(SO$_4$)$_{3}{\cdot}$8H$_2$O samples. 
The DLs of the developed method for U and Th were estimated to be 0.04 mBq (U)/kg and 0.01 mBq (Th)/kg. 
These values are precision enough for our requirements, which are 5 mBq (U)/kg and 0.05 mBq (Th)/kg. 
The concentrations of U and Th in the three purified samples were estimated to be $<~0.04$ mBq (U)/kg and 0.06 $\pm$ 0.01 mBq (Th)/kg for sample A, 0.05 $\pm$ 0.01 mBq (U)/kg and 0.12 $\pm$ 0.02 mBq  (Th)/kg for sample B, and 0.64 $\pm$ 0.10 mBq (U)/kg and 1.77 $\pm$ 0.27 mBq (Th)/kg for sample C.

\section*{Acknowledgment}

This work was supported by the JSPS KAKENHI Grants Grant-in-Aid for Scientific Research on Innovative Areas No. 26104007, Grant-in-Aid for Specially Promoted Research No. 26000003, and Grant-in-Aid for Young Scientists No. 17K14290. 
We also thank the companies who provided the purified Gd$_2$(SO$_4$)$_{3}{\cdot}$8H$_2$O samples.

\end{document}